\documentclass[preprint,showpacs,preprintnumbers,amsmath,amssymb,natbib]{revtex4}

\usepackage{graphicx}
\usepackage{epsfig}		
\usepackage{dcolumn}
\usepackage{bm}
\def\0{\mbox{\tiny $0$}}
\def\1{\mbox{\tiny $1$}}
\def\2{\mbox{\tiny $2$}}
\def\3{\mbox{\tiny $3$}}
\def\4{\mbox{\tiny $4$}}
\def\5{\mbox{\tiny $5$}}
\def\6{\mbox{\tiny $6$}}
\def\7{\mbox{\tiny $7$}}
\def\8{\mbox{\tiny $8$}}
\def\9{\mbox{\tiny $9$}}

\def\f14{\mbox{\tiny $\frac{1}{4}$}}

\def\bb#1{\mbox{\footnotesize $(#1)$}}
\def\bbb#1{\mbox{\footnotesize $\left(#1\right)$}}
\renewcommand{\baselinestretch}{1.4}

\begin{document}

\title{The Hamiltonian formalism for scalar fields coupled to gravity in a cosmological background}

\author{A. E. Bernardini}
\email{alexeb@ufscar.br}
\altaffiliation[On leave of absence from]{~Departamento de F\'{\i}sica, Universidade Federal de S\~ao Carlos, PO Box 676, 13565-905, S\~ao Carlos, SP, Brasil.}
\author{O. Bertolami}
\email{orfeu.bertolami@fc.up.pt}
\altaffiliation[Also at~]{Instituto de Plasmas e Fus\~ao Nuclear, Instituto Superior T\'ecnico, Av. Rovisco Pais, 1, 1049-001, Lisboa.} 
\affiliation{Departamento de F\'isica e Astronomia, Faculdade de Ci\^{e}ncias da
Universidade do Porto, Rua do Campo Alegre 687, 4169-007, Porto, Portugal.}
\date{\today}
\renewcommand{\baselinestretch}{1.3}

\begin{abstract}
A novel routine to investigate the scalar fields in a cosmological context is discussed in the framework of the Hamiltonian formalism.
Starting from the Einstein-Hilbert action coupled to a Lagrangian density that contains two components - one corresponding to a scalar field Lagrangian, ${\mathcal{L}}_{\phi}$, and another that depends on the scale parameter, ${\mathcal{L}}_{a}$ - one can identify a generalized Hamiltonian density from which first-order dynamical equations can be obtained.
This set up corresponds to the dynamics of Friedmann-Robertson-Walker models in the presence of homogeneous fields embedded into a generalized cosmological background fluid in a system that evolves all together isentropically.
Once the generalized Hamiltonian density is properly defined, the constraints on the gravity-matter-field system are straightforwardly obtained through the first-order Hamilton
equations.
The procedure is illustrated for three examples of cosmological interest for studies of the dark sector: real scalar fields, tachyonic fields and generalized Born-Infeld tachyonic fields.
The inclusion of some isentropic fluid component into the Friedmann equation allows for identifying an exact correspondence between the dark sector underlying scalar field and an ordinary real scalar field dynamics.
As a final issue, the Hamiltonian formulation is used to set the first-order dynamical equations through which one obtains the exact analytical description of the cosmological evolution of a  generalized Chaplygin gas (GCG) with dustlike matter, radiation or curvature contributions.
Model stability in terms of the square of the sound velocity, $c_{s}^{2}$, cosmic acceleration, $q$, and conditions for inflation are discussed.
\end{abstract}

\date{\today}
\maketitle
\renewcommand{\baselinestretch}{1.4}

\section{Introduction}

Scalar fields are of utmost importance in theoretical physics.
The recent confirmation of the existence of the Higgs boson \cite{Higgs,Higgs2} revives the discussion that repeatedly appears in the literature about the elementarity of scalar fields (see e. g. Ref.~\cite{Hill} and \cite{Djouadi}).
In cosmology, scalar fields play a key role in inflation \cite{Olive} and, for this purpose, cannot be replaced by, for instance, vector fields \cite{Bento}.

The recent discovery that the Universe is expanding in an accelerated way \cite{Perl}, can be accommodated by a scalar field with features, that likewise for inflation, yield a negative pressure (see e. g. Ref.~\cite{Rev01} for a review).
Thus, scalar field appears in many cosmological proposals for dark energy: quintessence \cite{Zla98,Wan99,Ste99}, Chaplygin gas model \cite{Kam02,Bil02,Ber02}, $k$-essence \cite{Arm01}, etc.

Of course, the dark energy features accounted by these scalar fields can be instead due to a cosmological constant with a suitable value.
This has its own problems and a variable cosmological term was proposed in this context \cite{Ber00,Wet87,Pee87,Rat87}.
Scalar fields can also play the role of dark matter (see e. g. Refs.~\cite{Rog01,Rog02}, and refs. therein)

In this context, it is thus quite relevant to seek for new methods to study scalar fields in a cosmological context.
In this work, a novel routine to study scalar fields in a cosmological set up is presented using the Hamiltonian formalism extended to the gravitational sector.
This might include the possibility of tackling the coupling of scalar fields so that only their coupled evolution is isentropic.
This situation is found in models of interacting dark energy and dark matter \cite{20,21,22}, a discussion which has several common features with the treatment of the cosmon field \cite{Wet11}
and models with dynamical masses \cite{Mot04,Bro06A,alex,alex2}.

Our starting point is the Arnowitt-Deser-Misner (ADM) formalism, which is extended to encompass the coupling to scalar fields in the context of Friedmann-Robertson-Walker (FRW) models.
We develop a procedure for obtaining the first-order Hamilton equations for scalar fields coupled to gravity and embedded in a cosmological background.
It is shown how first-order equations of motion are generically and systematically obtained for 
any family of Lagrangians, so that the dynamics can be explicitly obtained from a Hamilton-Jacobi type equation.

Once the Hamilton equations of motion are obtained, the physical interpretation of quantities involved follows straightforwardly as in other branches of physics.
The relevant canonical variables are identified, providing, for instance, a natural set up for quantization.
The advantages of the Hamiltonian formulation lie not only in its use as an analytical tool, but also in the deeper insight it unfolds.
The form invariance of the Hamilton equations that we shall construct here is not constrained to specific classes or families of field Lagrangians and can be applied to a fairly general class of scalar fields.

For instance, our approach can be used to obtain analytical solutions for the widely studied hypothesis of unified dark energy and dark matter of the generalized Chaplygin gas (GCG) \cite{Kam02,Bil02,Ber02} modified by isentropic components.
The GCG has been shown to be consistent with the constraints from CMB, supernova, gravitational lensing surveys, and gamma ray bursts \cite{Ber03,Sup1,Ber04,Ber05,Ber03B,Ber06B}.
Considering additional isentropic components to the pure GCG model is a fairly natural way to encompass also radiation, matter and curvature contributions.
An interesting application of our first-order Hamiltonian equations is to adapt it to the real scalar field underlying the GCG model.
The resulting formulation allows for tracking the behavior of the equation of state parameter, $\omega$, the model stability through the squared speed of sound, $c_{s}^{2}$, and the cosmic acceleration, $q$.

The outline of this manuscript is as follows.
In section II we review the ADM formalism as to ensure the consistency of our construction with the Hamiltonian formalism of FRW models.
In section III, we obtain the Hamilton equations for scalar fields.
Our emphasis is on the description of the dynamics through first-order partial differential equations from which a Hamilton-Jacobi type equation can be established.
The constraints and the equations of motion are obtained for three relevant examples: real scalar fields, tachyonic fields and tachyonic-type fields derived from the generalized Born-Infeld Lagrangian \cite{Ber02}.
The inclusion of a fluid component with an explicit dependence on the scale parameter, $a$, into the Lagrangian system is considered in section IV.
An extension of the Hamiltonian definition results into the invariance of the Hamilton equations of motion such that first-order equations for an overall isentropically modified dark sector are obtained.
We point out that the inclusion of such an additional fluid component into the Lagrangian allows one to identify a systematic correspondence between a real scalar field and other non-canonical scalar fields.
This is particularly useful for the treatment of the dark sector and it allows, for instance, one to parameterize the dynamics of tachyonic fields through an auxiliary real scalar field Lagrangian.
In section V we review the main features of the GCG so to describe it with our Hamiltonian formalism in the presence of isentropic components.
We consider the underlying scalar field solution and address the issue of stability of linear perturbations.
This extension is referred to as modified GCG (mGCG).
Finally, we draw our conclusions in section VI.
 
\section{Theoretical preliminaries}

In order to settle the framework of our discussion, we review here the Arnowitt-Deser-Misner (ADM) formalism for gravity coupled to a scalar field.
We start with the Einstein-Hilbert action,
\begin{equation}
\mathcal{S} = \frac{1}{4\kappa}\int{d^{4}x\, \sqrt{-g} \mathcal{R}},
\label{HJ000}
\end{equation}
where $g = \det{(g_{\mu\nu})}$, and $\mathcal{R} = R^{\mu\nu}\,g_{\mu\nu}$ is the scalar curvature, $\kappa = 4 \pi G$ and we use the natural system of units ($c = \hbar = 1$).
The variation with respect to the metric, $g_{\mu\nu}$, yields the second-order Einstein field equations of motion.
Alternatively, the ADM formalism can describe the dynamics through its Hamiltonian formulation. The fields are therefore solved explicitly for the time derivatives in a kind of first-order formalism \cite{Misner}.

The suitable three-dimensional quantities to describe general relativity \cite{Misner} are therefore
\begin{equation}
g_{ij},
\quad \pi_{ij} = \sqrt{-g}\left(\Gamma_{kl}^0 - g_{kl}\,\Gamma_{mn}^0\, g^{mn}\right)g^{ik}g^{jl},
\quad N = (-g^{00})^{-1/2},
\quad \mbox{and}
\quad N_{i} = g_{0i},
\label{HJ000A}
\end{equation}
where the connection, $\Gamma_{ij}^{k}$, is regarded as an independent quantity, with {\em latin} indices running from $1$ to $3$.

This procedure provides the essential ingredients to include matter and fields into a Lagrangian density, $\mathcal{L}_{m}$.
The theory can be, for instance, specified for the most general form of an $SO(4)$-invariant metric in a $M = \mathbb{R} \times S^3$ topology \cite{BertolamiMourao}.
In this case, the homogeneity and the isotropy at the background level can be depicted from the line element of the Robertson-Walker (RW) metric,
\begin{equation}
ds^{2} = - N(t)^{2}\,dt^{2} + a(t)^2\,\left[\frac{dr^2}{1- k r^2} + r^2\,d\Omega^{2}\right] ,
\label{HJ002}
\end{equation}
with $k = \pm1,\,0$ corresponding to $S^3$, $H^3$ and $\mathbb{R}^3$ hypersurfaces, $\sqrt{-g} = N(t) \, a(t)^3$, where we have introduced the lapse function, $N(t)$, and the scale parameter, $a(t)$, as arbitrary non-vanishing functions of time, $t$, and $d\Omega^{2} = d\theta^2 + \sin{(\theta)}^2  d\phi^2$.

For a Lagrangian density independent of the space coordinates, for fields in the RW spacetime the integration over these coordinates yields the volume of the hypersurface (for $a = 1$) as a multiplicative factor $V$ \cite{BertolamiMourao}.
Thus, the relevant action is
\begin{eqnarray}
\mathcal{S}_{eff}(t') = \frac{1}{V} \mathcal{S}(t') &=& 
\int_{t'}{dt\, \sqrt{-g} \left[\frac{1}{4\kappa}\mathcal{R} + \mathcal{L}_{m}\right]},
\label{HJ002A}
\end{eqnarray}
where $\mathcal{L}_{m} = \mathcal{L}_{a} +\mathcal{L}_{\phi}$ denotes the Lagrangian components corresponding to a perfect background fluid, $\mathcal{L}_{a}$, and to an homogeneous and isotropic scalar field with Lagrangian, $\mathcal{L}_{\phi}$, where $\phi$ can be associated to the dark sector.

For the RW metric, the Einstein-Hilbert Lagrangian can be expressed in terms of the components of the Ricci tensor and the scalar curvature. For $k = 0$,
\begin{equation}
R_{00} = 3\left[\frac{\dot{a}}{a}\frac{\dot{N}}{N} - \frac{\ddot{a}}{a}\right] = - 3\left[N\frac{d}{dt}\left(\frac{H}{N}\right) + H^2\right],
\label{HJ003A}
\end{equation}
and
\begin{equation}
\mathcal{R} = \frac{6}{N^2}\left[\left(\frac{\dot{a}}{a}\right)^2 - \frac{\dot{a}}{a}\frac{\dot{N}}{N} + \frac{\ddot{a}}{a}\right] = 
\frac{6}{N^2}\left[N\frac{d}{dt}\left(\frac{H}{N}\right) + 2 H^2\right],
\label{HJ003B}
\end{equation}
where the {\em dot} corresponds to derivatives with respect to $t$, and we have introduced the Hubble expansion rate,
\begin{equation}
H = \frac{\dot{a}}{a},
\label{HJ003E}
\end{equation}
such that
\begin{equation}
G_{00} = R_{00} - \frac{1}{2} g_{00} \mathcal{R} = 3 H^2,
\label{HJ003C}
\end{equation}
since $g_{00} = -N^{2}$.

The above results lead to 
\begin{equation}
\frac{1}{4\kappa} \sqrt{-g} \mathcal{R} = \frac{3}{2\kappa}\left( \dot{\omega} - \frac{a^3}{N}H^2\right),
\label{HJ00D}
\end{equation}
where we have introduced an auxiliary variable, $\omega(t) = a(t)^3\, H(t)/ N(t)$, such that its corresponding Lagrangian component proportional to $\dot{\omega}$ can be absorbed by the time integration of the effective action.
Thus, the effective action becomes
\begin{eqnarray}
\mathcal{S}_{eff}(t') &=& 
\int_{t'}{dt\, \left[\frac{3}{2\kappa}\frac{a^3}{N}H^2 + N\,a^3 \,\left(\tilde{\mathcal{L}}_{a} +\tilde{\mathcal{L}}_{\phi}\right)\right]},
\label{HJ002C}
\end{eqnarray}
and the so-called Hamiltonian constraint \cite{Misner} is obtained by considering the variation of $\mathcal{S}_{eff}$ with respect to the lapse function, $N$, in the following way,
\begin{eqnarray}
-\frac{\partial}{\partial N}\mathcal{L}_{eff}(t') &=& 
\frac{\partial}{\partial N} \left[\frac{3}{2\kappa}\frac{a^3}{N}H^2 - N\, \left(\tilde{\mathcal{L}}_{a} +\tilde{\mathcal{L}}_{\phi}\right)\right] = 0 ,
\label{HJ002D}
\end{eqnarray}
where we have introduced the transformed Lagrangian density, $\tilde{\mathcal{L}}_{a,\phi} = \sqrt{g^{(3)}} \mathcal{L}_{a,\phi}$, with $\sqrt{g^{(3)}} = a^{3}$, suitable to match the dynamics of the FRW universe.
The above equation results into 
\begin{equation}
-\frac{3}{2\kappa}\frac{a^3}{N^2}H^2 = \tilde{\mathcal{L}}_{m} + N\,\frac{\partial \tilde{\mathcal{L}}_{m}}{\partial N}.
\label{HJ002E}
\end{equation}

There exists several cases of cosmological relevance for which the stress-energy tensor is described by
\begin{equation}
T^{a,\phi}_{\mu\nu} = -2 \frac{\partial \mathcal{L}_{a,\phi}}{\partial g^{\mu\nu}} +   g_{\mu\nu} \mathcal{L}_{a,\phi},
\label{HJ007}
\end{equation}
from which follows that
\begin{eqnarray}
- N^{-2}\,T^{a,\phi}_{00} 
&=&
N^{-2}\left( 2 \frac{d N}{d g^{00}} \frac{\partial \mathcal{L}_{a,\phi}}{\partial N} - g_{00} \mathcal{L}_{a,\phi}\right) \nonumber\\
&=& N\, \frac{\partial \mathcal{L}_{a,\phi}}{\partial N} + \mathcal{L}_{a,\phi},
\label{HJ007}
\end{eqnarray} 
and, as previously done for $\mathcal{L}$ with respect to $\tilde{\mathcal{L}}$, introducing $\tilde{T}_{00} = \sqrt{g^{(3)}}T_{00}$, Eq.~(\ref{HJ002E}) becomes
\begin{equation}
\frac{3}{2\kappa}\frac{a^3}{N^2}H^2 = \frac{1}{N^2}\tilde{T}^{m}_{00} \quad\Leftrightarrow\quad \frac{3}{2\kappa}H^2 = T^{m}_{00},
\label{HJ007A}
\end{equation}
and hence the constraint Eq.~(\ref{HJ002E}) corresponds to the Friedmann's equation.

Notice that variation with respect to the scale factor leads to the equation of motion of which Friedmann's equation is the first integral of motion.
Likewise, the variation with respect to $\phi$ yields the field equation of motion.
The Hamiltonian density can then be identified with the usual energy density related to the Hubble expansion rate through
\begin{equation}
\mathcal{H}_{\phi} = \rho_{\phi} = \frac{3}{2\kappa} H^{2},
\label{HJ008}
\end{equation}
so that the two formalisms
\begin{equation}
\mathcal{H}_{\phi} \rightarrow \tilde{\mathcal{H}}_{\phi} \qquad \Leftrightarrow \qquad\mathcal{L}_{\phi} \rightarrow \tilde{\mathcal{L}}_{\phi}
\label{HJ008A}
\end{equation}
are equivalent to each other.

In the following sections we shall extend the analysis to include an isentropic fluid with energy density $\rho_a \equiv \rho(a)$.
In the process, a generalized Hamiltonian, $\mathcal{K}_{\phi}$, is introduced as,
\begin{equation}
\mathcal{H}_{\phi} =  \frac{3}{2\kappa}H^2 \qquad \rightarrow \qquad \mathcal{K}_{\phi} =  \frac{3}{2\kappa}H^2 - \rho_a,
\label{HJ008B}
\end{equation}
that leads to the correspondence,
\begin{equation}
\mathcal{K}_{\phi} \rightarrow \tilde{\mathcal{K}}_{\phi} \qquad \Leftrightarrow \qquad\mathcal{L}_{\phi} \rightarrow \tilde{\mathcal{L}}_{\phi}.
\label{HJ008C}
\end{equation}
such that Hamilton equations remain invariant.

\section{Hamiltonian equations for scalar fields}

In this section we address the issue of encompassing scalar fields which are relevant for cosmology, and dark energy modeling in particular.
We shall concretely consider an homogeneous real scalar field and the tachyon field.
Let one then supposes that the full action of a scalar field theory is given by the Einstein-Hilbert term plus the scalar field Lagrangian density denoted by $\mathcal{L}_{\phi} \equiv \mathcal{L}(\phi,\,\dot{\phi})$, as
\begin{equation}
\mathcal{S}_{eff} = \int{dt\, \sqrt{-g} \left[\frac{1}{4} \mathcal{R} + \mathcal{L}_{\phi} \right]},
\label{HJ001}
\end{equation}
where we have set $\kappa = 4\pi G$ equals to unity henceforward.
To obtain the Hamiltonian equations for the scalar field, $\phi$, we shall recover the transformed Lagrangian density, $\tilde{\mathcal{L}}_{\phi} = \sqrt{g^{(3)}} \mathcal{L}_{\phi}$, such as in the previous section.
Conjugate to the field component, $\phi$, the canonical momentum density, $\tilde{\pi}_{\phi}$, is given by
\begin{equation}
\tilde{\pi}_{\phi} = \frac{\partial \tilde{\mathcal{L}}_{\phi}}{\partial \dot{\phi}}.
\label{HJ004}
\end{equation}
Such that the independent quantities are now $\phi = \phi(t)$ and $\tilde{\pi}_{\phi} \equiv \tilde{\pi}_{\phi}(t)$, which define the phase-space describing the dynamical evolution of the classical field.
The natural generalization to the Hamiltonian density, $\tilde{\mathcal{H}}$, is given by
\begin{equation}
\tilde{\mathcal{H}_{\phi}} = \dot{\phi} \tilde{\pi}_{\phi} - \tilde{\mathcal{L}}_{\phi},
\label{HJ004A}
\end{equation}
from which the Hamiltonian equations are straightforwardly obtained as
\begin{eqnarray}
\frac{\partial \tilde{\mathcal{H}}_{\phi}}{\partial \tilde{\pi}_{\phi}} &=&
\dot{\phi}
            + \tilde{\pi}_{\phi} \frac{\partial \dot{\phi}}{\partial \tilde{\pi}_{\phi}}
            -  \frac{\partial \tilde{\mathcal{L}}_{\phi}}{\partial \tilde{\pi}_{\phi}} \nonumber\\
&=& \dot{\phi} + \left[\tilde{\pi}_{\phi}  - \frac{\partial \tilde{\mathcal{L}}_{\phi}}{\partial \dot{\phi}} \right] \frac{\partial \dot{\phi}}{\partial \tilde{\pi}_{\phi}},
\label{HJ004A1}
\end{eqnarray}
and
\begin{eqnarray}
\frac{\partial \tilde{\mathcal{H}}_{\phi}}{\partial \phi} &=&
\left[\tilde{\pi}_{\phi}  - \frac{\partial \tilde{\mathcal{L}}_{\phi}}{\partial \dot{\phi}} \right] \frac{\partial \dot{\phi}}{\partial \phi}
        - \frac{\partial \tilde{\mathcal{L}}_{\phi}}{\partial \phi}.
\label{HJ004A2}
\end{eqnarray}
By observing that the expression within brackets is null, then
\begin{eqnarray}
\frac{\partial \tilde{\mathcal{H}}_{\phi}}{\partial \tilde{\pi}_{\phi}} &=&
\dot{\phi},
\label{HJ004A1B}
\end{eqnarray}
and
\begin{eqnarray}
\frac{\partial \tilde{\mathcal{H}}_{\phi}}{\partial \phi} &=& - \frac{\partial \tilde{\mathcal{L}}_{\phi}}{\partial \phi}
= - \partial_{\mu} \left(\frac{\partial \tilde{\mathcal{L}}_{\phi}}{\partial (\partial_{\mu}\phi)}\right)
=
- \dot{\tilde{\pi}}_{\phi} - \partial_{j} \left(\frac{\partial \tilde{\mathcal{L}}_{\phi}}{\partial (\partial_{j}\phi)}\right),
\label{HJ004A2B}
\end{eqnarray}
where, in the last step, we have used the Euler-Lagrange equation and, for completeness, we have included a spatial dependence (index $j$) for the scalar field.
Since the scalar field is homogeneous, Eq.~(\ref{HJ004A2}) simplifies to
\begin{eqnarray}
\frac{\partial \tilde{\mathcal{H}}_{\phi}}{\partial \phi} &=&
- \dot{\tilde{\pi}}_{\phi},
\label{HJ004A2BB}
\end{eqnarray}
from which one can observe that
\begin{eqnarray}
\dot{\tilde{\mathcal{H}}}_{\phi}
&=&
\frac{\partial \tilde{\mathcal{H}}_{\phi}}{\partial \phi} \dot{\phi}
+\frac{\partial \tilde{\mathcal{H}}_{\phi}}{\partial \tilde{\pi}_{\phi}} \dot{\tilde{\pi}}_{\phi}
+\frac{\partial \tilde{\mathcal{H}}_{\phi}}{\partial t} \nonumber\\
&=&
- \dot{\tilde{\pi}}_{\phi} \dot{\phi} + \dot{\phi} \dot{\tilde{\pi}}_{\phi} + \frac{\partial \tilde{\mathcal{H}}_{\phi}}{\partial t} \nonumber\\
&=& \frac{\partial \tilde{\mathcal{H}}_{\phi}}{\partial t}
\label{HJ005A1}
\end{eqnarray}
can be re-written as
\begin{equation}
\frac{\partial \tilde{\mathcal{H}}_{\phi}}{\partial t} = \frac{\partial \tilde{\mathcal{L}_{\phi}}}{\partial t} = - \left(\frac{d}{dt} \sqrt{g^{(3)}}\right) \mathcal{L}_{\phi} = -(3 H \sqrt{g^{(3)}}) \mathcal{L}_{\phi} \neq 0,
\label{HJ005A2}
\end{equation}
since we have assumed that $\tilde{\pi}_{\phi}$ and $\phi$ are the canonically conjugated independent variables.

Getting back to the re-scaled Lagrangian, $\mathcal{L}_{\phi} = \tilde{\mathcal{L}}_{\phi} /\sqrt{g^{(3)}}$, and defining a re-scaled conjugate momentum, $\pi_{\phi} \equiv \partial \mathcal{L}_{\phi}/\partial \dot{\phi}$, the above equations can be simplified.
Introducing the re-scaled Hamiltonian density as $\mathcal{H}_{\phi} = \tilde{\mathcal{H}}_{\phi} / \sqrt{g^{(3)}}$, then
\begin{equation}
\mathcal{H}_{\phi} = \pi_{\phi} \dot{\phi} - \mathcal{L}_{\phi},
\label{HJ005}
\end{equation}
and the first Hamilton equation of motion follows
\begin{equation}
\frac{\partial \mathcal{H}_{\phi}}{\partial \pi_{\phi}}  = \dot{\phi} \equiv \frac{\partial \tilde{\mathcal{H}}_{\phi}}{\partial \tilde{\pi}_{\phi}},
\label{HJ005A5}
\end{equation}
which coincides with Eq.~(\ref{HJ004A1B}).
To obtain the other Hamilton equation, we substitute the result from Eq.~(\ref{HJ005A2}) into the time derivative of $\mathcal{H}$,
\begin{equation}
\dot{\mathcal{H}}_{\phi} = \frac{1}{\sqrt{g^{(3)}}} \dot{\tilde{\mathcal{H}}}_{\phi} - 3 H \mathcal{H}_{\phi},
\label{HJ005A35}
\end{equation}
for which we obtain
\begin{equation}
\dot{\mathcal{H}}_{\phi} = - 3 H (\mathcal{H}_{\phi} -\mathcal{L}) = - 3 H \pi_{\phi} \dot{\phi}.
\label{HJ005A3}
\end{equation}

In the absence of explicit dependence on $g$ (and consequently on $a(t)$ and on $t$) then $\partial \mathcal{H} /\partial t$ is null, such that
\begin{eqnarray}
\dot{\mathcal{H}}_{\phi} &=&
\frac{\partial \mathcal{H}_{\phi}}{\partial \phi} \dot{\phi}
+
\frac{\partial \mathcal{H}_{\phi}}{\partial \pi_{\phi}}
\dot{\pi}_{\phi}.
\label{HJ005A4}
\end{eqnarray}

Equating this to the right side of Eq.~(\ref{HJ005A3}) and using Eq.~(\ref{HJ005A5}) one easily obtains
the second Hamilton equation,
\begin{equation}
\frac{\partial \mathcal{H}_{\phi}}{\partial \phi}  = - \dot{\pi}_{\phi}  - 3 H \pi_{\phi}.
\label{HJ005A7}
\end{equation}

To sum up, for a given Lagrangian density, $\mathcal{L}_{\phi}$, one can easily find the conjugate momentum $\pi_{\phi}$, such that the Hamiltonian density can be obtained and the corresponding equations of motion can be established through Eqs.~(\ref{HJ005A5}) and (\ref{HJ005A7}).
Of course, the coupling with gravity implies that this procedure must be considered in the context of the $3+1$ ADM formalism.

Turning to an application of the above equations, the Hamilton-Jacobi ordinary constraint is essentially expressed in terms of $\pi_{\phi} \equiv \pi(\phi)$.
This results into $\mathcal{H}_{\phi} \equiv \mathcal{H}(\phi)$ and $\dot{\mathcal{H}}_{\phi} = (d\mathcal{H}_{\phi}/d\phi) \,\dot{\phi}$, that is
\begin{equation}
\dot{\phi} = \frac{\partial \mathcal{H}_{\phi}}{\partial \pi_{\phi}},
\label{HJ0061}
\end{equation}
and from Eq.(\ref{HJ005A3}) rewritten as
\begin{equation}
\frac{d\mathcal{H}_{\phi}}{d\phi} = - 3 H \pi_{\phi},
\label{HJ0062}
\end{equation}
into the further simplified form,
\begin{equation}
\frac{d\mathcal{H}_{\phi}}{d\phi} = 3 H \frac{d H}{d\phi} ~~~~ \rightarrow ~~~~ \pi_{\phi}
= - \frac{d H}{d\phi}.
\label{HJ009}
\end{equation}

From this point, the first-order equations of motion for any class of scalar fields are
Eqs.~(\ref{HJ0061}) and (\ref{HJ009}).
One can summarize this Hamiltonian formalism as follows: {\em i)} One computes the conjugate momentum, $\pi_{\phi}$, and through a Hamilton-Jacobi type equation one can set $\pi_{\phi} \equiv \pi(\phi)$; {\em ii)} Eq.~(\ref{HJ0061}) gives $\phi(t)$; {\em iii)} Eq.~(\ref{HJ009}) yields $a(t)$ and $\phi(a)$.
Essentially, the Hamiltonian equations correspond to two coupled first-order equations for $\phi(t)$ and $H(\phi)$.
To exemplify the applicability of this technique, we shall consider three cases of interest, given their use in the modeling of dark energy: real scalar fields, tachyonic fields and tachyonic fields arising from the generalized Born-Infeld action \cite{Ber02}.

\subsection{The real scalar field Lagrangian}

The equation of motion for a real scalar field, $\chi$, is obtained from a Lagrangian density,
\begin{equation}
\mathcal{L}_{\chi} = \frac{1}{2}\partial_{\mu} \chi \partial^{\mu} \chi - V(\chi),
\label{HJ0001}
\end{equation}
that allows for computing the stress-energy tensor from which one obtains the  energy density and pressure:
\begin{equation}
\rho_{\chi} =  \frac{1}{2}\dot{\chi}^{2} + V(\chi) \equiv \mathcal{H}_{\chi},
\label{HJ0001B}
\end{equation}
\begin{equation}
p_{\chi} =  \frac{1}{2}\dot{\chi}^{2} - V(\chi).
\label{HJ0001C}
\end{equation}
The conjugate momentum of the theory is easily obtained as
\begin{equation}
\pi_{\chi} \equiv \frac{\partial \mathcal{L}_{\chi}}{\partial \dot{\chi}} = \dot{\chi},
\label{HJ0001D}
\end{equation}
which corresponds to Eq.~(\ref{HJ0061}).
The Hamilton-Jacobi constraint is assumed in order to set
\begin{equation}
\mathcal{H}_{\chi} \equiv \mathcal{H}(\chi) = \frac{1}{2}\pi(\chi)^2 +V(\chi),
\label{HJ0001F}
\end{equation}
such that the explicit dependence on $\chi$ is evinced.
Eq~(\ref{HJ009}) combined with Eq.~(\ref{HJ0061}) yields
\begin{equation}
\dot{\chi} =  - \frac{1}{3 H} \frac{d\mathcal{H}_{\chi}}{d\chi} =  -\frac{d H}{d\chi},
\label{HJ0001E}
\end{equation}
that corresponds to the Hamiltonian formulation for the real scalar field problem.
Naturally the obtained solution is the same to the one obtained from the Euler-Lagrange equation of motion.
In fact, setting $\pi_{\chi}= \dot{\chi}$ into Eq.~(\ref{HJ005A3}), one easily obtains the usual second-order equation of motion for the problem,
\begin{equation}
\ddot{\chi} + 3 H \dot{\chi} + \frac{d V}{d\chi} = 0,
\label{HJ0001G}
\end{equation}
that shows the equivalence between the Hamiltonian and the Lagrangian formulations.

\subsection{The tachyonic field Lagrangian}

The equation of motion for a tachyonic field, $\varphi$, is obtained from a Lagrangian density given by
\begin{equation}
\mathcal{L}_{\varphi} = -  V(\varphi) (1 - \partial_{\mu} \varphi \partial^{\mu} \varphi)^{\frac{1}{2}},
\label{HJ0002}
\end{equation}
from which, as previously specified, one obtains the energy density and pressure,
\begin{equation}
\rho_{\varphi} = V(\varphi) (1 - \dot{\varphi}^{2})^{-\frac{1}{2}} \equiv \mathcal{H}_{\varphi},
\label{HJ0002B}
\end{equation}
\begin{equation}
p_{\varphi} =  -V(\varphi) (1 - \dot{\varphi}^{2})^{\frac{1}{2}}.
\label{HJ0002C}
\end{equation}
Once again, the conjugate momentum of the theory is obtained as
\begin{equation}
\pi_{\varphi} \equiv \frac{\partial \mathcal{L}_{\varphi}}{\partial \dot{\varphi}}
= V(\varphi) \dot{\varphi} (1 - \dot{\varphi}^{2})^{-\frac{1}{2}}
\label{HJ0002D}
\end{equation}
that leads to
\begin{equation}
\dot{\varphi} = \frac{\pi_{\varphi}}{\left(\pi_{\varphi}^{2} + V(\varphi)^{2}\right)^{\frac{1}{2}}} =\frac{\pi_{\varphi}}{\mathcal{H}_{\varphi}}
\label{HJ0002D1}
\end{equation}
where
\begin{equation}
\mathcal{H}_{\varphi} \equiv \mathcal{H}(\varphi) = \left[\pi(\varphi)^{2} + V(\varphi)^{2}\right]^{\frac{1}{2}}.
\end{equation}
The Hamilton-Jacobi equation follows as the potential has an explicit dependence on $\varphi$.
From Eqs.~(\ref{HJ0062}) and (\ref{HJ0002D1}), one also obtains
\begin{equation}
\dot{\varphi} =  - \frac{1}{3 H} \frac{d\mathcal{H}_{\varphi}}{d\varphi} \frac{1}{\mathcal{H}}_{\varphi} =  -\frac{2}{3 H^{2}}
\frac{d H}{d\varphi},
\label{HJ0002E}
\end{equation}
that corresponds to an alternative formulation along the lines discussed above for the tachyonic field.
From Eq.~(\ref{HJ005A3}), substituting $\pi_{\varphi}$ from Eq.~(\ref{HJ0002D}), one obtains the known second-order equation of motion for the problem,
\begin{equation}
\ddot{\varphi} + (1 - \dot{\varphi}^{2}) \left(3 H \dot{\varphi} + \frac{1}{V} \frac{d V}{d\varphi}\right) = 0,
\label{HJ0002G}
\end{equation}
ensuring the equivalence between the Hamiltonian and the Lagrangian formulations.

\subsection{Tachyonic fields arising from the generalized Born-Infeld Lagrangian density}

Finally, let us consider the equation of motion for the tachyonic-type field, $\theta$, arising from the generalized Born-Infeld Lagrangian density \cite{Ber02},
\begin{equation}
\mathcal{L}_{\theta} = -  V(\theta)
\left[1 - (\partial_{\mu} \theta \partial^{\mu} \theta)^{\frac{1+\alpha}{2\alpha}}\right]^{\frac{\alpha}{1+\alpha}},
\label{HJ0003}
\end{equation}
which clearly reproduces the tachyonic Born-Infeld Lagrangian for $\alpha = 1$.
This action is associated to the generalized Chaplygin gas (GCG) model \cite{Ber02}, a possible candidate for unification of dark energy and dark matter. 
The energy density and pressure are given respectively by
\begin{equation}
\rho_{\theta} = V(\theta) \left[1 - \dot{\theta}^{\frac{1+\alpha}{\alpha}}\right]^{-\frac{1}{1+\alpha}}\equiv \mathcal{H}_{\theta},
\label{HJ0003B}
\end{equation}
and
\begin{equation}
p_{\theta} =  -V(\theta) \left[1 - \dot{\theta}^{\frac{1+\alpha}{\alpha}}\right]^{\frac{\alpha}{1 + \alpha}}.
\label{HJ0003C}
\end{equation}
The conjugate momentum of the theory is straightforwardly given by
\begin{equation}
\pi_{\theta} \equiv \frac{\partial \mathcal{L}_{\theta}}{\partial \dot{\theta}}
= V(\theta) \dot{\theta}^{\frac{1}{\alpha}} \left[1 - \dot{\theta}^{\frac{1+\alpha}{\alpha}}\right]^{-\frac{1}{1+\alpha}},
\label{HJ0003D}
\end{equation}
that leads to
\begin{equation}
\dot{\theta}
= \frac{\pi_{\theta}^{\alpha}}{\left[\pi_{\theta}^{1 + \alpha} + V(\theta)^{1 + \alpha}\right]^{\frac{\alpha}{1 + \alpha}}}
= \left(\frac{\pi_{\theta}}{\mathcal{H}_{\theta}}\right)^{\alpha},
\label{HJ0003D1}
\end{equation}
for which, again the constraint
\begin{equation}
\mathcal{H}_{\theta} \equiv \mathcal{H}(\theta) = \left[\pi(\theta)^{1 + \alpha} + V(\theta)^{1 + \alpha}\right]^{\frac{1}{1 + \alpha}},
\end{equation}
is imposed.
From Eqs.~(\ref{HJ0062}) and (\ref{HJ0003D1}), one obtains
\begin{equation}
\dot{\theta} =  \left(- \frac{1}{3 H} \frac{d\mathcal{H}_{\theta}}{d\theta} \frac{1}{\mathcal{H}_{\theta}}\right)^{\alpha} =
\left(-\frac{2}{3 H^{2}} \frac{d H}{d\theta} \right)^{\alpha},
\label{HJ0003E}
\end{equation}
which is admissible only for $d H/ d\theta < 0$.
Following the previous arguments for the real scalar and the tachyonic fields, when the potential has its explicit dependence on $\theta$ written in terms of
\begin{equation}
V(\theta) = \left[\mathcal{H}_{\theta}^{1+\alpha} - \pi_{\theta}^{1+\alpha}\right]^{\frac{1}{1+\alpha}},
\label{HJ0003F}
\end{equation}
the above formulation based on Eq.~(\ref{HJ0003E}) is equivalent to the second-order equation of motion for the problem.
Indeed, substituting $\pi_{\theta}$ from Eq.~(\ref{HJ0003D}) into Eq.~(\ref{HJ005A3}), one obtains
\begin{equation}
\ddot{\theta} + \alpha (\dot{\theta}^{\frac{\alpha-1}{\alpha}} - \dot{\theta}^{2}) \left(3 H \dot{\theta}^{\frac{1}{\alpha}} + \frac{1}{V} \frac{d V}{d\theta}\right) = 0.
\label{HJ0003G}
\end{equation}

To sum up, the described Hamiltonian formulation of the problem is completely implemented when one is able to set the dependence of the Hamiltonian $\mathcal{H}(\phi,\,\pi_{\phi})$ in terms of the underlying scalar field, $\phi$.
This is achieved through the Hamilton-Jacobi equation, which is given either explicitly by $\pi_{\phi} = \pi(\phi)$(or $\mathcal{H}_{\phi} = \mathcal{H}(\phi)$), or implicitly through some equation of state for $\phi$, for which we have given three examples.

As we have anticipated in the introduction, the treatment of specific solutions is beyond the scope of the present discussion.

\section{Coupling to a fluid component and the correspondence involving real scalar fields}

The results from section II allow for the formulation of the cosmological problem in a simple form as they correspond to a systematic extension of the Hamilton-Jacobi formalism to homogeneous and isotropic cosmologies.

Let us consider the dynamical evolution of a dark sector in the presence of an additional fluid component into the Friedmann equation for the flat space case,
\begin{equation}
H^2 = \left(\frac{\dot{a}}{a}\right)^2 = \frac{2}{3} (\rho_{\phi} + \rho_{a}).
\label{HJ0001M}
\end{equation}

In fact, a more flexible constraint can be considered in agreement with the First Law of Thermodynamics, $d\rho = \mu dn + n \,T ds$.
The equation of state can be specified by the function $\rho(n, s)$, i.e., the energy density as a function of number density and entropy per particle.
The quantity $\mu = \partial\rho/\partial n = (\rho_{a} + p_{a})/n$ is the chemical potential.
This leads to
\begin{equation}
\frac{d\rho_{a}}{da} - \frac{1}{n}\frac{dn}{da}(\rho_{a} + p_{a}) = \frac{d\rho_{a}}{da} + \frac{3}{a}(\rho_{a} + p_{a}) = n\,T \frac{ds}{da},
\end{equation}
where it is assumed that $n\propto a^{-3}$.
Non-isentropic fluids, with $ds/da\neq 0$, require the inclusion of a compensating component into the Lagrangian density (c. f. Eq.~(\ref{HJ005A2})). 
This results into an extra dynamical term into the second Hamilton's equation, Eq~(\ref{HJ005A7}), that can account for the non-isentropic effect due to $ds/da \neq 0$.
A treatment that covers many cases of interest \cite{Wet87,Wet11} is the extended isentropic scenario
\begin{equation}
\sum_{m = \phi, a}\left({\frac{d\rho_{m}}{da} - \frac{1}{n}\frac{dn}{da}(\rho_{m} + p_{m})}\right) = 0,
\label{HJ010}
\end{equation}
where the entropy variation is compensated by the field dynamics or some mass varying mechanism.

In the usual isentropic scenario for a fluid or field, the stress-energy tensor is of a perfect fluid
\begin{equation}
T_{\mu\nu} = \rho U_{\mu}U_{\nu} + p(g_{\mu\nu} +U_{\mu}U_{\nu}),
\end{equation}
where the pressure is defined as $p = n \partial\rho/\partial n - \rho$, from which follows that $ds/da = 0$.
In addition, when imposing the stress-energy tensor covariant conservation, the perfect fluid also implies the covariant conservation of particle number, $n$. 
Thus, from the above point of view, there is no particular reason to regard the choice of the on-shell Lagrangian density $\mathcal{L}_{a} = p_{a}$ as preferable over $\mathcal{L}_{a} = -\rho_{a} + \mu\, n$.
This degeneracy of the Lagrangian density of a perfect fluid is inherent of general relativity, but can be lifted in more general theories (see Ref.~\cite{Lobo} and refs. therein).

Of course, the background fluid is assumed to be homogeneously and isotropically distributed into the Universe and, therefore, can be identified with the cosmic background radiation ($\rho_{a} \propto a^{-4}$), with the cosmological background of non-relativistic particles ($\rho_{a} \propto a^{-3}$), or with some curvature component ($\rho_{a} \propto a^{-2}$).

The Hamilton equations of motion are generalized to include a transformed Hamiltonian density, $\mathcal{K}_{\phi} \equiv \mathcal{K}(\phi,\,\pi_{\phi})$, given by
\begin{equation}
\mathcal{K}_{\phi} = \frac{3}{2} H^2 - \rho_a(a) = \rho_{\phi} \equiv \rho(\phi,\, \dot{\phi}) \equiv \rho(\phi,\, \pi_{\phi}),
\label{HJ012}
\end{equation}
i. e. the transformed Hamiltonian density, $\mathcal{K}$, is identified only with the energy density of the field sector, and not with the total energy density.

By following the same procedure described in the previous section, one obtains that
\begin{equation}
\mathcal{K}_{\phi} = \pi_{\phi} \dot{\phi} - \mathcal{L}_{\phi},
\end{equation}
so that the generalized Hamilton equations are extensions of Eqs.~(\ref{HJ0061}) and (\ref{HJ0062}),
\begin{equation}
\dot{\phi} = \frac{\partial \mathcal{K}_{\phi}}{\partial \pi_{\phi}},
\label{HJ0061B}
\end{equation}
and
\begin{equation}
\frac{d\mathcal{K}_{\phi}}{d\phi} = - 3 H \pi_{\phi},
\label{HJ0062B}
\end{equation}
obtained from
\begin{equation}
\dot{\mathcal{K}}_{\phi} = - 3 H \pi_{\phi} \dot{\phi}.
\label{HJ0062BK}
\end{equation}

After some manipulation involving the overall isentropic condition, Eq.~(\ref{HJ010}), it is easy to verify that the substitution of Eq.~(\ref{HJ012}) into Eq.~(\ref{HJ0062B}) leads to the usual dynamical constraint for FRW models,
\begin{equation}
\dot{H} = - (\rho_{a} + p_{a}) - \pi_{\phi} \dot{\phi} = - (\rho_{a} + p_{a}) - (\mathcal{K}_{\phi} + \mathcal{L}_{\phi}),
\label{HJ0062C}
\end{equation}
i. e. the Raychaydhuri-type equation,
\begin{equation}
\dot{H} = - (\rho_{a} + p_{a} + \rho_{\phi} + p_{\phi}),
\label{HJ0062C2}
\end{equation}
follows when the Lagrangian, $\mathcal{L}_{\phi}$, is identified with $p_{\phi}$.

At this point, one can recover the previous results by replacing the Hamiltonian density, $\mathcal{H}_{\phi}$, by the modified one, $\mathcal{K}_{\phi}$, into Eqs.~(\ref{HJ0001}) - (\ref{HJ0003G}).

Although the Hamilton-Jacobi constraint can be introduced directly into Eqs.~(\ref{HJ0061B}) and (\ref{HJ0062B}), for $\phi = \chi,\, \varphi,$ and $\theta$, the first-order equations can also be written as
\begin{equation}
\dot{\chi} =  - \frac{1}{3 H} \frac{d\mathcal{K}_{\chi}}{d\chi},
\label{HJ0001EK}
\end{equation}
for real scalar fields; as
\begin{equation}
\dot{\varphi} =  - \frac{1}{3 H} \frac{d\mathcal{K}_{\varphi}}{d\varphi} \frac{1}{\mathcal{K}_{\varphi}},
\label{HJ0002EK}
\end{equation}
for tachyonic fields; and as
\begin{equation}
\dot{\theta} =  \left(- \frac{1}{3 H} \frac{d\mathcal{K}_{\theta}}{d\theta} \frac{1}{\mathcal{K}_{\theta}}\right)^{\alpha},
\label{HJ0003EK}
\end{equation}
for tachyonic-type fields arising from the generalized Born-Infeld Lagrangian density.

The constraints over the relevant potentials follow the same substitution of $\mathcal{H}$ by $\mathcal{K}$.
Obviously, the dependence of $\rho_{a}$ on $a$, as well as the analytical form of the potential are necessary for obtaining explicit solutions of the above equations, for instance, as those discussed in Ref.~\cite{Bas2000}. 

\subsection{The correspondence involving real scalar fields}

Let us finally make the point that the inclusion of an isentropic fluid introduces an additional degree of freedom into the Friedmann equation.
The Hamiltonian formulation allows one to systematically identify a correspondence between the real scalar field and some other class of field.
This is established when the degree of freedom introduced by the isentropic fluid energy density, $\rho_a(a)$, is transferred into an {\em auxiliary} real scalar field potential, $U(\psi)$, through the mathematical correspondence
\begin{equation}
\rho_a(a) = - U(\psi),
\label{HJ020}
\end{equation}
with $\psi$, dynamically constrained by $\psi = \sqrt{3} \ln{(a)}$, which results into a Hubble rate parameterized by
\begin{equation}
H = \frac{\dot{a}}{a} = \frac{1}{\sqrt{3}} \dot{\psi}.
\label{HJ021}
\end{equation}
By replacing the above results into the expression for the modified Hamiltonian density, Eq.~(\ref{HJ012}), one can identify the following correspondence between $\psi$ and a generic field $\phi$ as 
\begin{equation}
\mathcal{K}_{\phi}= \rho(\phi,\, \pi_{\phi}) ~~\leftrightarrow~~ \mathcal{K}_{\psi} = \rho(\psi,\, \pi_{\psi}),
\label{HJ022}
\end{equation}
since, from Eqs.(\ref{HJ020}) and (\ref{HJ021}), one has
\begin{equation}
\rho(\phi,\, \pi_{\phi}) = \frac{3}{2} H^2 - \rho_a(a) =  \frac{1}{2}\dot{\psi}^2 + U(\psi) = \rho(\psi,\, \pi_{\psi}),
\label{HJ022}
\end{equation}
with $\pi_{\psi} = \dot{\psi}$.

The Hamilton equations than follow
\begin{eqnarray}
\frac{\partial\mathcal{K}_{\psi}}{\partial\pi_{\psi}} &=& \dot{\psi}, \nonumber\\
\frac{\partial\mathcal{K}_{\psi}}{\partial\psi} &=& \frac{dU}{d\psi} = -\dot{\pi}_{\psi} - 3 H \pi_{\psi},
\label{HJ023B}
\end{eqnarray}
that lead to the well-known equation of motion for real scalar fields,
\begin{equation}
\dot{\pi}_{\psi} + 3 H \pi_{\psi} +  \frac{dU}{d\psi}  = 0.
\label{HJ023C}
\end{equation}

The explicit analytical dependence of the potential $U(\psi)$ on $\psi$ is naturally obtained from the equation of state for the isentropic fluid, $\rho_{a} \equiv \rho_a(p_a)$, and vice-versa.
In the same way, Eq.~(\ref{HJ023C}) represents the constraint given by the isentropic fluid condition from Eq.~(\ref{HJ010}).

As we have verified through Eq.~(\ref{HJ023B}), the above correspondence keeps unchanged the generalized Hamilton equations of motion for any scalar field.
As an example, let us consider the dynamics of the dark sector driven by a tachyonic field in a cosmological model with a positive curvature.
In this case, the Friedman equation for the dark sector is modified by an extra energy density component given by $\rho_{a}(a) = -1/a^{2}$, which obeys the isentropic fluid condition.
By identifying the dependence of the scale factor, $a$, with the {\em auxiliary} real scalar field, $\psi$,
\begin{eqnarray}
a\equiv a(\psi) =\exp{\left(-\frac{\psi}{\sqrt{3}}\right)},
\label{HJ026}
\end{eqnarray}
one obtains
\begin{equation}
\rho(\varphi,\, \pi_{\varphi}) = V(\varphi)(\pi_{\varphi}^{2} + 1 )^{\frac{1}{2}} \equiv
\rho(\psi,\, \pi_{\psi})   = \frac{1}{2} \pi_{\psi}^{2} + \exp{\left(-\frac{2\psi}{\sqrt{3}}\right)},
\label{HJ026A}
\end{equation}
and, considering that both fields, $\varphi$, and the {\em auxiliary} one, $\psi$, follows the same equation of state,
\begin{equation}
p(\varphi,\, \pi_{\varphi}) = - V(\varphi)(\pi_{\varphi}^{2} + 1 )^{-\frac{1}{2}} \equiv
p(\psi,\, \pi_{\psi})   = \frac{1}{2} \pi_{\psi}^{2} - \exp{\left(-\frac{2\psi}{\sqrt{3}}\right)},
\label{HJ026B}
\end{equation}
their correspondence results into to the following transformations,
\begin{equation}
\pi_{\psi} = \sqrt{V(\varphi)} \frac{\pi_{\varphi}}{(\pi_{\varphi}^{2} + 1)^{\frac{1}{4}}},
\label{HJ027A}
\end{equation}
and
\begin{equation}
\psi = -\frac{\sqrt{3}}{2}\ln{\left[
\frac{V(\varphi)}{2}\frac{\pi_{\varphi}^{2} + 2}{(\pi_{\varphi}^{2} + 1)^{\frac{1}{2}}}
\right]} \label{HJ027B}.
\end{equation}
When, for instance, $V(\varphi) = 1$ , the explicit dependence of $\pi_{\varphi}$ on the scale factor is given by
\begin{equation}
\pi_{\varphi} = \frac{\sqrt{2}}{a^{2}}(1 - a^{4})^{\frac{1}{4}}
\left[(1 - a^{4})^{\frac{1}{2}} + 1 \right]^{\frac{1}{2}},
\label{HJ027AA}
\end{equation}
which essentially illustrates the discussed correspondence.
Of course, one could admit extensions that involve more complex expressions for $V(\varphi)$, as well as other correspondences involving either tachyonic-type fields or a cosmic background with $\rho \propto a^{w}$, with $w$ arbitrary.

\section{Exact solution for the GCG with additional isentropic components}

The GCG model is described by the equation of state \cite{Kam02} where the fluid pressure is given by,
\begin{equation}
p = - A_{s} \rho_{0} \left(\frac{\rho_{0}}{\rho}\right)^{\alpha},
\label{gcg001}
\end{equation}
which can be deduced from a generalized Born-Infeld action \cite{Ber02} and reproduced through a canonical real scalar field.
This model interpolates the dust dominated Universe in the past, where $\rho\propto a^{-3}$, and a de-Sitter phase at late times, where $\rho\propto -p$.
This evolution is phenomenologically constrained by the parameters $\alpha$ and $A_{s}$, both positive, and with $0 < \alpha \leq 1$.

Once introduced into the Einstein equation, the above equation of state can be described in terms of an underlying real scalar field.
In this case, the background fluid energy density and pressure are given Eqs.~(\ref{HJ0001B}) and (\ref{HJ0001C}) such that the equation of motion is written as Eq.~(\ref{HJ0001G}).
Introducing Eq.~(\ref{gcg001}) into Eq.~(\ref{HJ0001G}), one obtains the analytical solution for $\chi$ as
\begin{equation}
\chi\bb{a} = - \frac{1}{\sqrt{6} (\alpha + 1)}\ln{\left[\frac{\sqrt{1 - A_{s}(1 - a^{3(\alpha + 1)})} - \sqrt{1 - A_{s}}}{\sqrt{1 - A_{s}(1 - a^{3(\alpha + 1)})} + \sqrt{1 - A_{s}}}\right]},
\label{gcg24}
\end{equation}
with
\begin{equation}
\chi_{0} \equiv \chi\bb{a_{0} = 1} = - \frac{1}{\sqrt{6} (\alpha + 1)}\ln{\left[\frac{1 - \sqrt{1 - A_{s}}}{1 + \sqrt{1 - A_{s}}}\right]}.
\label{gcg25}
\end{equation}
from which one readily finds the scalar field potential,
\begin{equation}
V\bb{\chi} =
\frac{1}{2}A_{s}^{\frac{1}{\alpha + 1}}\rho_{0}
\left\{
\left[\cosh{\left(\sqrt{\frac{3}{2}}(\alpha + 1) \chi\right)}\right]^{\frac{2}{\alpha + 1}}
+
\left[\cosh{\left(\sqrt{\frac{3}{2}} (\alpha + 1) \chi\right)}\right]^{-\frac{2\alpha}{\alpha + 1}}
\right\}.
\label{gcg26}
\end{equation}
These above solutions can be obtained from Ref.~\cite{Ber02} (where $8\pi G/3$ was set equal to unity).

Let us now get the above results from Hamilton-Jacobi description of the real scalar field $\chi$.
If one sets
\begin{equation}
H = \frac{\dot{a}}{a}= y\bb{\chi},
\label{xx}
\end{equation}
as a first-order differential equation for the scale factor, one notices from Eq.~(\ref{HJ0001E}) that
\begin{equation}
\dot{\chi} = - \frac{d y}{d\chi} \equiv y_{,\chi},
\label{xxx}
\end{equation}
i. e. a first-order differential equation for the real scalar field.
As pointed out in the previous section, Eq.~(\ref{xxx}) depends on the nature of the field \cite{Bas2000}, i. e. on the starting Lagrangian from which Hamilton-Jacobi equations are deduced.
Eq.~(\ref{HJ0001G}) and the nature of the real scalar field identified by  Eqs.~(\ref{HJ0001B}) and (\ref{HJ0001C}) implies through Eqs.~(\ref{xx}) and (\ref{xxx}) that the potential has the simplified form
\begin{equation}
V\bb{\chi} = \frac{3}{2} y^{2} - \frac{1}{2} y^{2}_{,\chi},
\label{teste2}
\end{equation}
which prescribes the first-order formulation of the problem.

Turning back to the GCG solution, an elementary mathematical exercise allows one to identify the auxiliary function $y(\chi)$ with
\begin{equation}
y\bb{\chi} =
\left(A_s^{\frac{1}{\alpha + 1}}\rho_{0}\right)^{\frac{1}{2}}
\cosh{\left[\sqrt{\frac{3}{2}} (\alpha + 1) \chi \right]}^{\frac{1}{\alpha + 1}},
\label{teste1}
\end{equation}
so that through Eq.~(\ref{gcg26}) one would easily have obtained the expression for $V(\chi)$.

One shall verify in the following that the analytical form of Eqs.~(\ref{teste2}) and (\ref{teste1}) supports an initial {\em ansatz} giving by Eq.~(\ref{gcg24}) for obtaining the exact solution for the total energy density for the mGCG universe described by the GCG in the presence of an additional component of isentropic fluid.

\subsection{Inclusion of isentropic components}

Considering that the cosmic inventory is by the GCG with energy density, $\rho_{\chi}$, and the pressure, $p_{\chi}$, defined in terms of a real scalar field, $\chi$, the inclusion of an extra component of an isentropic fluid modifies the Friedmann equation, which can now be rewritten as
\begin{equation}
\frac{3}{2} H^{2} = \zeta\bb{a} = \left(\rho_{\chi} + \rho^0_{w}\sigma\bb{a}\right) = \left(\frac{1}{2}\dot{\chi}^{2} + V\bb{\chi} + \rho^0_{w}\sigma\bb{a}\right)
\label{gcg003A2}
\end{equation}
and, from Eq.~(\ref{HJ0062BK}),
\begin{equation}
\dot{K} = \dot{\zeta} - \rho^0_{w}\dot{\sigma} = -3 H \dot{\chi}^{2},
\label{gcg003B2}
\end{equation}
where we have introduced the auxiliary variables $\zeta\bb{a}$, which represents the total energy of the mGCG, and $\sigma\bb{a} = a^{-3(1 + w)}$, which represent the dynamical evolution of the energy density of the extra isentropic component.
In this case,  one should set $w = 0$ for pressureless non-relativistic matter, $w = 1/3$ for radiation, and $w = -1/3$ for a positive curvature contribution.
For other values of $w$ one could also have some exotic contribution.

By dividing the above equation by $\dot{\sigma}$ and observing that $\dot{\zeta}/\dot{\sigma} \equiv \zeta_{,\sigma}\equiv d\zeta/d\sigma$, one obtains
\begin{equation}
\zeta_{,\sigma} + \left(\frac{2 \dot{\chi}^2}{H \dot{\sigma}}\right) \, \zeta = \rho^0_{w},
\label{gcg003C2}
\end{equation}
from which, using Eq.~(\ref{gcg24}), one has
\begin{eqnarray}
\frac{2 \dot{\chi}^2}{H \dot{\sigma}} &=& \frac{2 a\, \chi_{,a}^{2}}{\sigma_{,a}} \nonumber\\
&=& -\frac{1}{1 + w} \frac{(1-A_{s}) a^{3(w-\alpha)}}{(1-A_{s}) a^{-3(\alpha+1)} + A_{s}}\nonumber\\
&=& -\frac{1}{1 + w} \frac{(1-A_{s}) \sigma^{\frac{\alpha-w}{w+1}}}{(1-A_{s}) \sigma^{\frac{\alpha+1}{w+1}} + A_{s}},
\label{gcg003D2}
\end{eqnarray}
resulting in the following simplified form of the first-order non-homogeneous differential equation
\begin{equation}
\zeta_{,\sigma} - \frac{1}{1 + w}
\frac{(1-A_{s}) \sigma^{\frac{\alpha-w}{w+1}}}
{(1-A_{s}) \sigma^{\frac{\alpha+1}{w+1}} + A_{s}}\zeta
= \rho^{0}_{w}.
\label{gcg003E2}
\end{equation}

Finally, the exact solution for the total energy density, $\rho\equiv\zeta\bb{a}$, of the mGCG is obtained as
\begin{eqnarray}
\zeta\bb{a} &=& \frac{3}{2} H^{2} =
\rho_{\chi}\bb{a} + \rho_{w}\bb{a} = \rho_{\chi}\bb{a} +\rho^{0}_{w}a^{-3(w + 1)}\nonumber\\
&=& \left(A_{s} +\frac{1-A_{s}}{a^{3(\alpha + 1)}}\right)^{\frac{1}{\alpha + 1}}
\left[\rho_{0} - \rho^{0}_{w} + \frac{\rho^{0}_{w}}{A_{s} a^{3(w + 1)}}\,
_2F_1\bbb{\frac{w + 1}{\alpha + 1},\,\frac{1}{\alpha + 1},\,\frac{\alpha + w + 2}{\alpha + 1},\,\frac{A_{s}-1}{A_{s}\,a^{3(\alpha + 1)}}} \right],
\label{gcg003F2}
\end{eqnarray}
where $_2F_1$ is the Gauss's hypergeometric function.

The explicit dependence of $\zeta\bb{a}$ on the scale factor, $a$, allows one to obtain the Hubble parameter and the pressure and energy density components of the cosmic inventory.

In the next section, we shall compare some relevant dynamical variables for both scenarios: GCG and mGCG, i. e.  the dark sector modified by an isentropic fluid.
To sum up, assuming the solution for the underlying scalar field of the original GCG as a constraint for Hamilton-Jacobi equation allows for extending the above results to the {\em tachyonic} formulation of the GCG since real scalar and {\em tachyonic} fields are shown to be related with each other.

\subsection{Dynamical properties, stability and conditions for inflation}

From this point on we shall illustrate the properties of the mGCG obtained from Eq.~(\ref{gcg003F2}) for the equation of state parameter, $\omega$, the cosmic acceleration, $q$, the squared speed of sound, $c_{s}^{2}$, and the Hamilton-Jacobi inflation condition depicted by the parameter $\epsilon$.
We shall consider the GCG in the presence of:
a {\em dustlike} matter component with $w = 0$ and $\rho^{0}_{w} = 0.04$,
a radiation component with $w = 1/3$ and $\rho^{0}_{w} = 0.0001$ (photons and three ultra-relativistic neutrino families), and
a curvature contribution with $w = -1/3$ and $\rho^{0}_{w} = 0.0001$.
In order to show the generality of the analytical results that we have obtained, we also include an exotic-radiation component with $w = 1/3$ and $\rho^{0}_{w} = 0.04$,
and a highly exotic-radiation component with $w = 1/2$ and $\rho^{0}_{w} = 0.04$.
We assume that $\rho_{0} = 1$ and that all densities are expressed in units of the critical density, $\rho_{Crit}$.

In addition, it is well-known that a striking feature of the GCG fluid is that its energy density interpolates between a dust dominated phase, $\rho_{\chi} \propto a^{-3}$, in the past, and a de-Sitter phase, $\rho_{\chi} = -p_{\chi}$, at late times.
It can be shown that the GCG model admits inhomogeneities and that, in particular, in the context of the Zeldovich approximation, these evolve in a qualitatively similar way as in the $\Lambda$CDM model \cite{Ber02,Ber05}.
This evolution is controlled by the model parameters, $\alpha$ and $A_{s}$.
The latter value arises from the matching of the model with supernova, CMB data, and cosmic topology \cite{Ber04,Ber06B}.
In particular, in what concerns $\alpha$, the most stringent observational constraints arise from WMAP5 and structure formation data that imply into $\alpha \lesssim 0.2$ \cite{Ber04B}.

\begin{figure}
\vspace{-2.5 cm}
\centerline{\psfig{file= 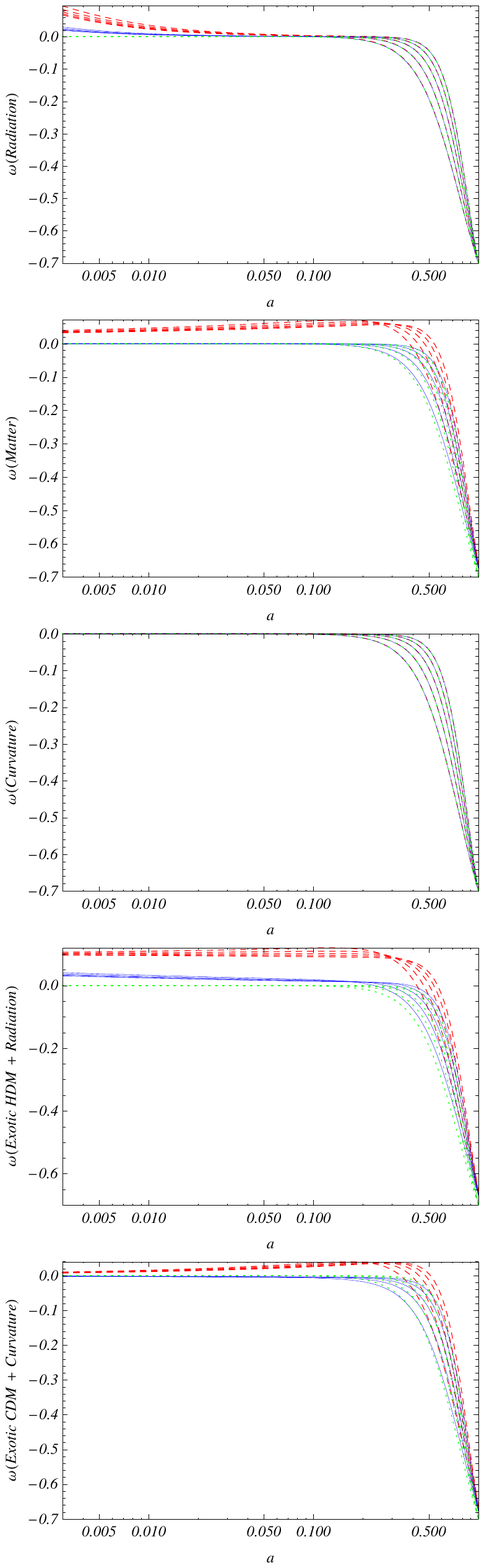, width= 15 cm}}
\vspace{-1.0 cm}
\caption{\footnotesize Equation of state, $\omega = p /\rho$, as function of the scale factor for the cosmological scenario driven by the mGCG (dashed red lines) for several choices of $\alpha$ (in the range between $0.1$ and $0.9$) and for $A_{s} = 0.7$.
The mGCG exact results are compared with those for the effective GCG (dotted green lines), and for the effective GCG perturbatively modified by an isentropic fluid (solid blue lines).
First and second pictures correspond to realistic modifications to the original GCG scenario where respectively radiation, with $\Omega_{\gamma} = 10^{-4}$ and $w_{\gamma} = 1/3$,
and cold (barionic) matter, with $\Omega_{m} = 0.04$ and $w_{m} = 0$, were separately included into the Friedmann equation.
Third picture corresponds to modifications due to the inclusion of an exotic curvature isentropic component, with $\Omega_{\kappa} = 10^{-4}$ and $w_{\kappa} = -1/3$.
Forth and fifth pictures are academic examples respectively corresponding to modifications due to the inclusion of
exotic HDM plus radiation, with $\Omega = 0.04$ and $w_{\gamma} = 1/10$, and exotic CDM plus curvature, with $\Omega = 0.04$ and $w_{\gamma} = -1/10$.}
\label{FigPap01}
\end{figure}

For the five abovementioned cosmological configurations, one can depict $\omega = p/\rho$ for the mGCG as shown in Fig.~\ref{FigPap01}.
The GCG parameters are chosen to cover a wide range of phenomenological possibilities.
Herein we have considered choices for $\alpha$ in the range from $0.1$ to $0.9$ and $A_{s} = 0.7$.
The mGCG exact results are compared with those for the effective GCG (dotted green lines), and for the effective GCG perturbatively modified by an isentropic fluid (solid blue lines).
From that it is easy to identify the modifications due to the inclusion of isentropic components of radiation and barionic matter.
As expected, curvature (cosmological constant phase component) does not change the results in a relevant way, and the inclusion of exotic components, in this case, is not relevant.

\begin{figure}
\vspace{-2.5 cm}
\centerline{\psfig{file= 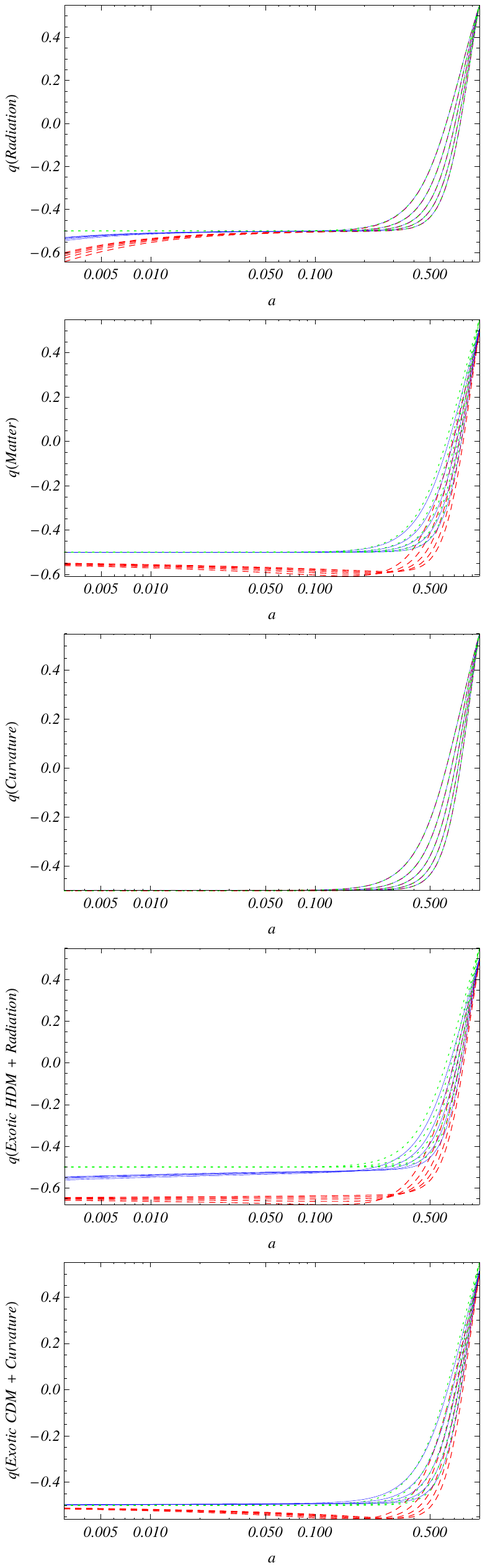, width= 17 cm}}
\vspace{-1.3 cm}
\caption{\footnotesize Cosmic acceleration $q = \ddot{a}/(a\,\rho)$, as function of the scale factor for a cosmological scenario driven by the mGCG (dashed red lines).
The ordering of the pictures, the values for the energy density of the isentropic fluid component, and the GCG parameters are in correspondence with those used in Fig.~\ref{FigPap01}.
Again the mGCG exact results are compared with those for the effective GCG (dotted green lines), and for the effective GCG perturbatively modified by an isentropic fluid (thin blue lines).}
\label{FigPap02}
\end{figure}

In Fig.~\ref{FigPap02} we depict the cosmological acceleration, $q$.
Notice that the inclusion of radiation, or more significatively of barionic matter, delays the onset for the accelerated expansion phase of the GCG.
One can also notice the discrepancy between the exact solution given by dashed red lines with respect to the perturbative solution given by solid blue lines.
Again, the inclusion of exotic isentropic fluids just amplifies the effects due to the inclusion of matter and radiation.

\begin{figure}
\vspace{-2.5 cm}
\centerline{\psfig{file= 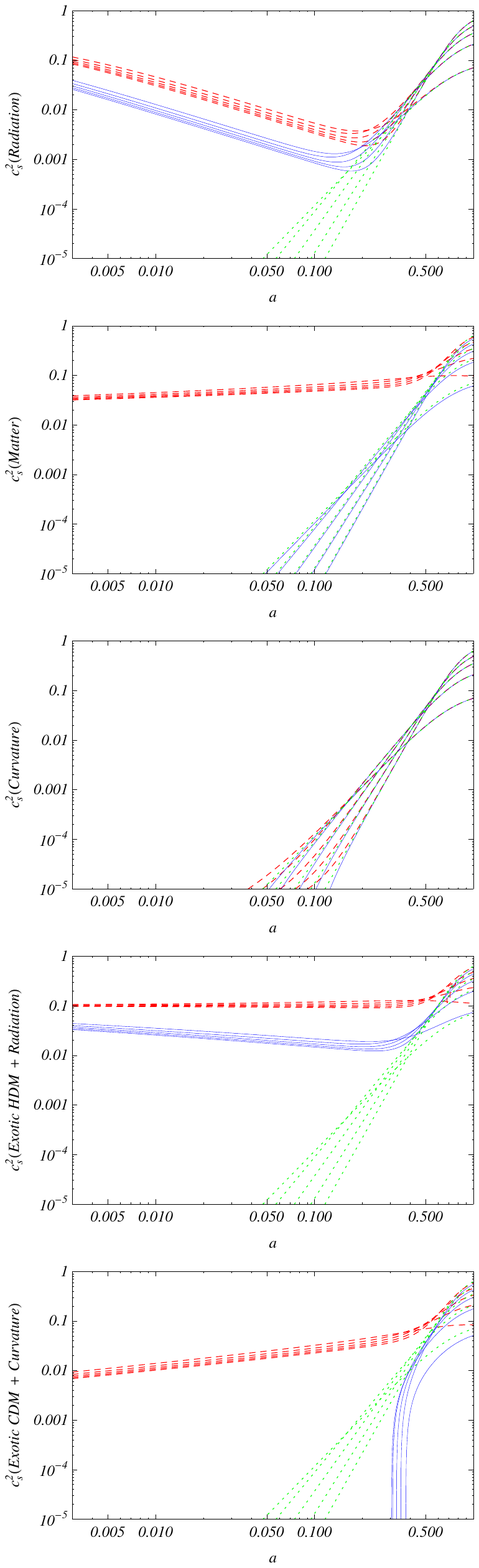, width= 17 cm}}
\vspace{-1.3 cm}
\caption{\footnotesize Squared speed of sound $c^{2}_{s} = \mbox{d}p/\mbox{d}\rho$, as function of the scale factor for a cosmological scenario driven by the mGCG (dashed red lines).
The comparative curves, the ordering of the pictures, the values for the energy density of the isentropic fluid component, and the GCG parameters are in correspondence with the previous figures (c. f.  Figs.~\ref{FigPap01} and \ref{FigPap02}).}
\label{FigPap03}
\end{figure}

Besides the conditions for the accelerated expansion of the Universe, in order to understand the possible range of values for $\alpha$, one has to consider the propagation of sound through the (m)GCG fluid, where sound velocity is given by $c_{s}^{2} = d p\bb{a}/ d\rho\bb{a}$.
A detailed quantitative analysis of the stability conditions for the GCG in terms of the squared speed of sound was discussed in Ref. \cite{Ber04,alex}.
Positive $c_{s}^{2}$ implies that $0 \leq \alpha \leq 1$.
The inclusion of isentropic components relaxes such constraints.
By computing the squared speed of sound one indeed verifies how a cosmological background isentropic fluid can substantially change the conditions for the propagation of linear perturbations.
As depicted in Fig.~\ref{FigPap03}, in spite of an exact convergence for late times, the difference between the mGCG and the GCG results are quite clear at early times, during the radiation dominated epoch.
Moreover, since the component of barionic matter is relatively large, as a fraction of the critical density, $\rho_m^0 \sim 0.04 \rho_{Crit}$, the perturbative formulation of the problem does not apply in a consistent way.

\begin{figure}
\centerline{\psfig{file= 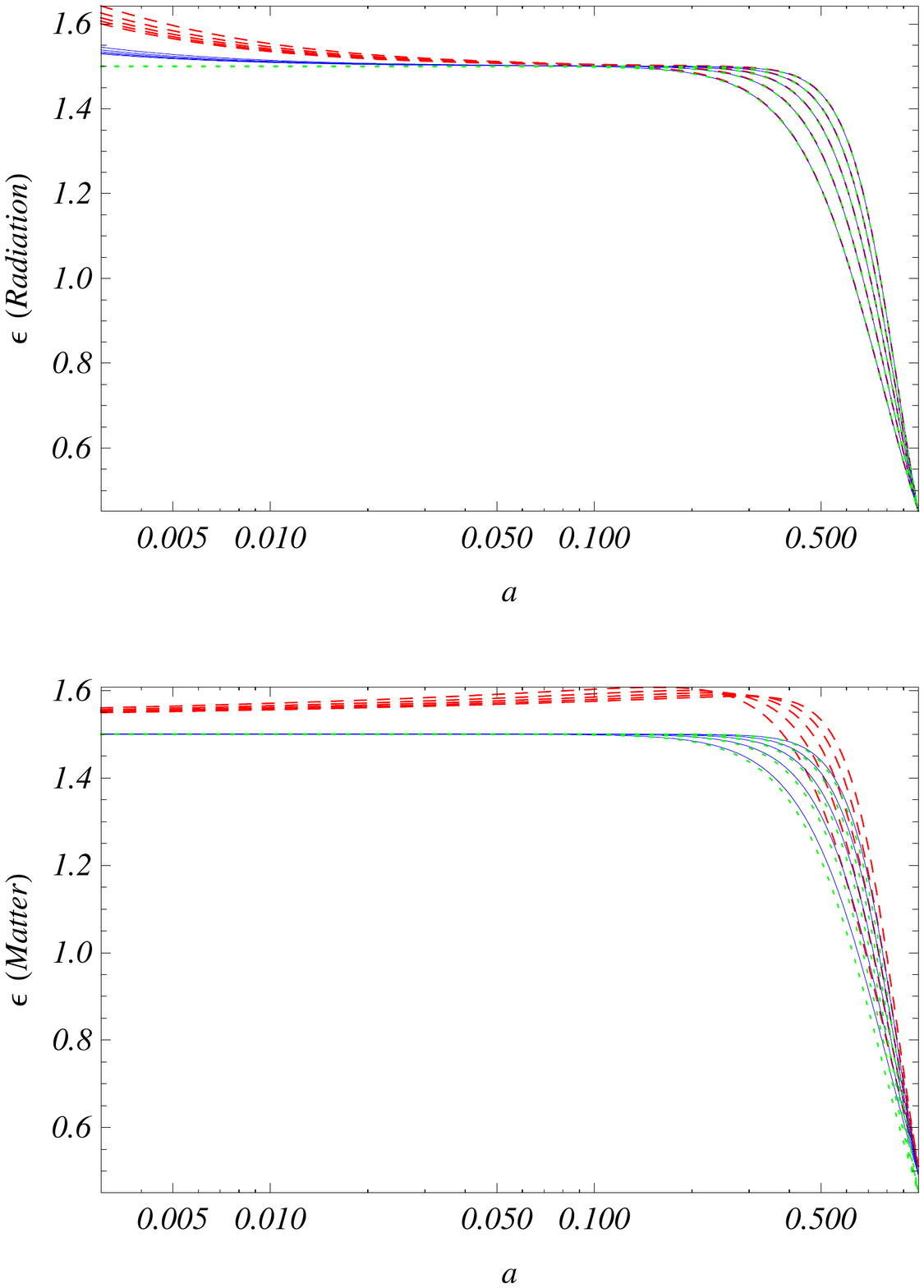, width= 13.0 cm}}
\caption{\footnotesize Hamilton-Jacobi inflationary parameter $\epsilon$, as function of the scale factor for a cosmological scenario driven by the mGCG (dashed red lines).
The mGCG exact results are compared with those for the effective GCG (dotted lines), and for the effective GCG perturbatively modified by an isentropic fluid (solid lines).
The first picture corresponds to modifications to the original GCG scenario where radiation, with $\Omega_{\gamma} = 10^{-4}$ and $w_{\gamma} = 1/3$, was include into the calculations.
The second picture corresponds to modifications to the original GCG scenario where cold (barionic) matter, with $\Omega_{m} = 0.04$ and $w_{m} = 0$, was included.}
\label{FigPap04}
\end{figure}

Finally, let us consider the inflationary slow-roll parameter $\epsilon$
\begin{equation}
\epsilon \equiv \epsilon(a) = -\frac{d\ln{H(a)}}{d\ln{a}}
\end{equation}
whose behavior is shown in Fig.~\ref{FigPap04}.
In the slow-roll regime, the condition for inflation is given by
\begin{equation}
(\mbox{Accelerated Expansion})~~ \ddot{a} > 0 ~~\Leftrightarrow~~\epsilon < 1,
\end{equation}
which is satisfied if one compares Fig.~\ref{FigPap04} with Fig.~\ref{FigPap02} for the region of accelerated expansion ($q > 0$).
Thus, our calculations are also consistent with the description of the onset of inflation.

\section{Conclusions}

In this work we have established a procedure for obtaining the first-order equations for the dynamics of scalar fields coupled to gravity.
The resulting Hamiltonian dynamics is more suitable for introducing Hamilton-Jacobi type equation in the context of FRW models in the presence of background matter.
Moreover, we have verified that the inclusion of an isentropic fluid component into the Friedmann equation allows for reproducing the effect of a generalized scalar field dynamics (for instance, the tachyonic one) through a correspondence to a real scalar field, without any additional constraints.

Our results are consistent with a similar analysis involving a kind of first-order formalism that considers some particular models described by scalar fields in generic spacetime \cite{Bas2000}.
The physical situations that we have exemplified in the paper concerned real scalar fields and tachyonic fields commonly used in models of dark energy.

As an example, the formalism was used to analyze the GCG, an isentropic fluid that interpolates the dust dominated phase in the past, where $\rho\propto a^{-3}$, and a cosmological constant de-Sitter phase at late times, where $\rho\propto -p$.

We have considered the possibility of enriching the GCG dynamics with isentropic fluid components, and obtained exact solutions of the mGCG model.
Our analysis show that the inclusion of an isentropic {\em well-behaved} component into the cosmic inventory plus the GCG is consistent with a positive squared speed of sound through out the cosmological evolution.

The Hamiltonian formulation discussed here provides the tools for adjusting the GCG parameters in the presence a cosmological fluid background as well as to enlarge the conditions for stability in the propagation of linear perturbations, compatible with the accelerated expansion of the Universe.

{\em Acknowledgments - The work of A. E. B. is supported by the Brazilian Agencies FAPESP (grant 12/03561-0) and CNPq (grant 300233/2010-8). The work of O. B. is partially supported by the Portuguese Funda\c{c}\~ao para Ci\^encia e Tecnologia (FCT) by the project PTDC/FIS/11132/2009}

\renewcommand{\baselinestretch}{1.0}
\scriptsize

\end{document}